\documentclass[sigconf,nonacm]{acmart}

\usepackage{graphicx}
\usepackage{xifthen}
\usepackage{versions}

\newcounter{mn}
\newcommand{\superscript}[1]{\ensuremath{{}^{\textrm{\scriptsize #1}}}}

\newcommand{\mntext}[1]{\colorbox{SkyBlue}{\begin{color}{black}#1\end{color}}}
\newcommand{\mn}[2][]{{\tiny\superscript{\mntext{\arabic{mn}}}}\marginpar{\scriptsize{
  \ifthenelse{\isempty{#1}}
  {\mntext{\parbox{0.95\marginparwidth}{\superscript{\arabic{mn}}~\raggedright{#2}}}}
  {\mntext{\parbox{0.95\marginparwidth}{\superscript{\arabic{mn}}#1 says: \raggedright{#2}}}}
}}\stepcounter{mn}}

\usepackage{color,subfig}
\usepackage{soul}

\newcommand{\anova}[5]{($\mathrm F_{#1,#2}=#3, p<#4, \eta_G^2=#5$)}
\newcommand{\p}[1]{($p<#1$)}
\newcommand{\peq}[1]{($p=#1$)}

\setcopyright{none}

\begin{document}

\date{}

\title{Augmented Unlocking Techniques for Smartphones Using Pre-Touch Information}

\fancyhf{} \fancyfoot[C]{\thepage}

\author{Matthew Lakier}
\email{mlakier@uwaterloo.ca}
\affiliation{University of Waterloo}

\author{Dimcho Karakashev}
\email{dzkaraka@uwaterloo.ca}
\affiliation{University of Waterloo}

\author{Yixin Wang}
\email{y3244wan@uwaterloo.ca}
\affiliation{University of Waterloo}

\author{Ian Goldberg}
\email{iang@uwaterloo.ca}
\affiliation{University of Waterloo}

\begin{abstract}
    Smartphones store a significant amount of personal and private information, and are playing an increasingly important role in people's lives. It is important for authentication techniques to be more resistant against two known attacks called shoulder surfing and smudge attacks. In this work, we propose a new technique called 3D Pattern. Our 3D Pattern technique takes advantage of a new input paradigm called pre-touch, which could soon allow smartphones to sense a user's finger position at some distance from the screen. We implement the technique and evaluate it in a pilot study (n=6) by comparing it to PIN and pattern locks. Our results show that although our prototype takes about 8 seconds to authenticate, it is immune to smudge attacks and promises to be more resistant to shoulder surfing.
\end{abstract}

\maketitle

\section{Introduction}

Smartphones are increasingly used to store private information such as personal photos, contacts, and financial information. However, smartphones are also frequently used in public spaces or in social gatherings, necessitating the protection of this private information via user authentication. Authentication or ``unlocking'' techniques include the common manual (e.g., PINs and gesture-based pattern locks) and biometric (e.g., fingerprint reading, iris scanning, and face recognition) techniques, and less commonly \textit{continuous authentication} techniques, which continuously monitor the user's behaviour, such as touch or swipe patterns, locking the device if it believes a different person has started to use it.

In this work, we focus on manual authentication techniques, because they are among the most common techniques used on smartphones as a way to protect private information. Even users utilizing fingerprint readers often are required to enter a PIN for added security, for example, when rebooting or authorizing payments.

Harbach et al.~\cite{HarbachAnatomy2016} showed that bystanders looking at other people's phones as they type their PIN (``shoulder surfers'') are able to reliably deduce the PINs. This unfortunately hinders the effectiveness of PINs. Many different PIN entry techniques have been proposed to improve shoulder-surfing resistance, such as scrambled keyboards~\cite{ScrambledKeyboardTan2005}, haptic and sound-based PINs~\cite{BianchiHaptic2011}, gestures (including swiping)~\cite{vonZezschwitzSwiPIN2015}, graphical PINs~\cite{GraphicalPassowordsChiang2013}, and techniques based on remembered user behaviour~\cite{UserPatternsTypingPasswordBuschek2015}. In general, techniques have a tradeoff between performance and resistance to shoulder surfing~\cite{HarbachAnatomy2016}.

Recently, there has been a move towards creating systems that support \emph{pre-touch sensing}, that is, using information about user's fingers just before the screen is actually touched~\cite{Xia0Latency2014,PreTouchHinckley2016,FingerIDKim2018}.
Similarly to how pre-touch information has been used for expanding target selection~\cite{PreTouchTargetSelYang2011}, we identify an opportunity to apply pre-touch sensing to improve the effectiveness of PIN entry techniques. We create a novel version of the Android pattern lock that expands the traditional $3\times3$ grid out of the screen into a $3\times3\times3$ cube. Points are connected by moving a finger in 3D space above the surface of the phone. Because pre-touch information is not available on current smartphones, we simulate pre-touch using a motion capture system. This enables us to implement a prototype version of the 3D Pattern lock.

A pilot study with six participants shows that the 3D Pattern technique is slower and more error-prone than the PIN and pattern techniques. These results could be partially attributed to the novelty of pre-touch input and the motion-capture approach used by the prototype, both of which would be ameliorated when pre-touch input becomes available in commonly used devices. We also find that the 3D Pattern technique has increased shoulder-surfing resistance compared to PINs. Further, the empirical CDF indicates that a larger study could reveal an improvement in shoulder-surfing resistance over the pattern technique as well. Finally, the 3D Pattern technique is naturally immune to smudge attacks~\cite{AvivSmudgeAttack2010}.

The main contributions of this paper are (1) the design of a novel smartphone authentication technique called 3D Pattern using pre-touch, and (2) an implementation and evaluation of the 3D Pattern technique in comparison to conventional PIN and pattern locks.

\section{Background and Related Work}

In this section, we discuss different types of attacks against smartphone authentication techniques, as well as past techniques designed to defend against these attacks.

\subsection{Shoulder Surfing}
Shoulder surfing is a widely known attack in which the adversary tries to infer the victim's authentication secret by looking over his or her shoulder. There is a significant body of research into mitigating the impact of shoulder-surfing attacks.
An in-depth survey conducted by Eiband et al.~\cite{EibandSurfingWild2017} considered the threat not only in the context of authentication, but also in the context of routine smartphone usage.
The survey showed that 130 out of 174 participants indicated that shoulder-surfing attacks occurred on public transportation. 
Victims most commonly defended against such an attack by modifying their posture or cancelling the authentication.
Furthermore, a study conducted by Harbach et al.~\cite{HarbachPerception2014} found the perceived risk of shoulder surfing to be high in only 11 of 3410 situations.
This demonstrates that people are not actively defending themselves against shoulder surfing, and more work is needed to improve the shoulder-surfing resistance of authentication techniques.

There is significant amount of previous work that has attempted to address shoulder surfing, but the proposed solutions either do not adequately defend against shoulder surfing or result in other problems such as longer authentication times or increased error rates.

\subsection{Smudge Attacks}
PIN keypads and pattern locks are commonly used methods for phone authentication. Unfortunately, these techniques are vulnerable to smudge attacks, because the user leaves oily residues on the screen. Previous work has demonstrated that smudge attacks are especially effective on pattern locks as users drag their fingers over the screen. Smudge attacks can also be used to limit the input space for PIN locks. Aviv et al.~\cite{AvivSmudgeAttack2010} found that as long as the line of sight is not perpendicular, it is easy to observe entered patterns based on smudges. Under ideal conditions, 92\% of the patterns entered were partially identifiable and 68\% of the entered patterns were fully recoverable. Under less ideal conditions, 37\% of the patterns were partially recoverable and 14\% were fully recoverable. 

These results demonstrate that even if the adversary is not able to actively observe the process of authentication, he or she can still recover the password with considerable success. In our work, we leverage pre-touch information to limit the number of touches the user makes on the screen, mitigating the effect of smudge attacks. 

\subsection{PIN and Password Locks}

Before the advent of smartphones, Tan et al.~\cite{TanMouseDrag2005} devised a password entry technique designed for a computer mouse; however, a similar method could be applied for touchscreens. In this technique, the user presses left and right to subtly highlight a letter on a scrambled keyboard. The user then uses the mouse to drag a tile on top of this letter. While dragging, the scrambled keyboard letters disappear so an onlooker cannot tell which letter is being selected.

Now that smartphones are commonplace, traditional authentication techniques have been adapted to work on the small touchscreens of smartphones. Kovelamudi et al.~\cite{KovelamudiScramble2016} compared speed and shoulder-surfing resistance of a scrambled PIN entry keypad and a normal PIN entry keypad. They found that the scrambled keypad was slower but more resistant to shoulder surfing.

Several works have examined the possibility of augmenting PIN keypads with gestures. SwiPIN, by von Zezschwitz et al.~\cite{vonZezschwitzSwiPIN2015}, divided the PIN keypad into two sections. Each number in each section corresponded to a different swipe gesture direction. Performing a swipe gesture on the correct section of the screen would insert the corresponding number. Their study demonstrated that this technique improved resistance against smudge attacks. Khan et al. introduced ``ForcePINs''~\cite{KhanForcePIN2018}, with which each PIN digit could be entered with different levels of finger pressure on the screen, to add an additional layer of challenge for shoulder surfers. However, results showed that there was no statistically significant difference in shoulder-surfing resistance between regular PINs and ForcePINs, because when users pressed harder, they also pressed for a noticeably longer time.

Other works have looked beyond purely visual representations of PINs by incorporating haptic and audio feedback. Bianchi et al.~\cite{BianchiHaptic2011} created an observation-resistant authentication technique by providing no visual clues to the user. The technique renders a wheel on the screen with identical sections. However, when users drag their fingers over the sections of the wheel, tactile feedback is presented with varying lengths and strengths. To select a section, users drag their fingers to the middle of the wheel. After each entry, the sections are shuffled to provide resistance against smudge attacks. Similarly, VibraInput~\cite{KuribaraRotateAlign2014} used an on-screen, rotary wheel with two levels. The outer level contained the letters A through D, each corresponding to a fixed vibration pattern (that has to be remembered by user). The inner level corresponded to the PIN numbers 0 through 9. Upon starting PIN entry, the phone would vibrate the pattern of a letter. The user would then rotate the outer wheel to align the letter with the number to select on the inner wheel. By repeating this process, the technique could use process of elimination to ascertain the PIN number. The overall technique would repeat until the entire PIN was entered.

Two-Thumbs-Up (TTU)~\cite{NyangTwoThumbs2018} prevents shoulder-surfing attacks by requiring the user to cover the screen with their hands. This forms a ``handshield'' and enters a challenge mode. If users move their hands away from the screen, the authentication technique disappears. TTU randomly associates five ``response'' letters with two digits each, presenting the digits and letters on either side of the screen. The user then has to tap on the letter corresponding to the next PIN digit. After a certain number (dependent on PIN length) of correctly selected letters, the authentication process is complete.

\subsection{Pattern Locks}
Harbach at al.~\cite{HarbachAnatomy2016} focused on comparing PIN locks and pattern locks. They were able to observe the behaviour of 134 smartphone users over one month, revealing differences between the two techniques. Results showed that although pattern locks are faster, users are six times as likely to make mistakes compared to PIN locks. When including failed attempts, there were no differences in authentication time between the two techniques.
When a user made a mistake entering a PIN or pattern, subsequent successful attempts took more time, presumably because the user took more care when repeating the authentication.
Visual feedback did not influence the error rate nor the entry time. Similarly, our 3D Pattern technique improves shoulder-surfing resistance by reducing visual feedback during authentication.

De Luca et al. suggested using a stroke-based visual authentication scheme~\cite{DeLucaPatternLock2007}, expecting visual patterns to be easier to remember in comparison to traditional numeric PINs or alphanumeric passwords. A similar technique, the pattern lock, was ultimately incorporated into the Android operating system. Unfortunately, as outlined above, pattern locks have been shown to be weak against shoulder surfing and smudge attacks. In contrast, DRAW-A-PIN, by Nguyen et al.~\cite{NguyenDrawDigit2017}, has the user draw each PIN digit on the screen using their finger. Results indicated that this approach was capable of mitigating shoulder surfing attacks.

\subsection{Graphical}

Another category of PIN entry techniques uses pictures or other graphics. In SemanticLock~\cite{OladePictures2018}, users arrange icons on the screen in a memorable way. The user is authenticated based on correct placement of the icons. In a similar work, Awase-E~\cite{TakadaAwaseE2003}, Takada and Koike leverage photos taken on a user's smartphone. The lock screen breaks a user-chosen photograph up into smaller chunks, and shows nine chunks of various photographs all at once. The user then has to select the tile from the correct photograph four times in a row to unlock the phone.

\subsection{Small-Scale Interaction}

Several works have attempted to mitigate the impact of smudge attacks by limiting the touch interaction to a small area on the screen. TinyLock~\cite{KwonSmudge2014} resists smudge attacks without trading off usability. Users draw their pattern in a tiny grid, making it harder to observe finger motions due to the small interaction area. To finish authentication, the user rotates a virtual wheel on top of the grid, distorting the smudges from the pattern. Similarly, ClickPattern~\cite{GuerarFakePattern2017} shows a keypad in a randomly shuffled order in a small area at the bottom of the screen. The user presses the keys to enter numbers corresponding to a pattern. This technique has the same lack of shoulder surfing resistance as the Android pattern lock because the pattern is visualized on the screen. However, it improves smudge attack resistance because of the small input area on the screen. Our 3D Pattern technique avoids the need to limit the interaction area by not requiring the user's finger to make contact with the screen.

\subsection{Behavioural Authentication}

There is considerable research exploring whether or not lock screens are even necessary at all, by applying \emph{continuous authentication}, also known as implicit authentication. Continuous authentication systems analyze an individual's regular patterns of touches on the screen, and build a model. A different user would have different patterns, and could be denied access by the system. With the Touchalytics project~\cite{FrankTouchalytics2013}, Frank et al. were able to use continuous authentication to identify the user with an error rate below 4\%, even after a week of elapsed time between training and testing.
Unfortunately, Khan et al.~\cite{KhanTargetedMimicry2016} showed that an attacker, merely watching a video of the target using their phone, could bypass swipe-based continuous authentication at least 75\% of the time.

Some work has explored applying the principles of continuous authentication to augment traditional lock screen techniques. Buschek et al.~\cite{UserPatternsTypingPasswordBuschek2015} use spatial touch features in addition to previously used temporal touch features on keyboards to verify users based on their individual text entry behaviours. Examples of spatial touch features include touch offsets, angles, and pressures. By incorporating such spatial features, user recognition accuracy was improved.

\subsection{Pre-Touch}

Many recent works on touchscreen interactions have started exploring pre-touch information; that is, positional information about the user's hands or fingers before making contact with the screen. For example, with TouchCuts and TouchZoom, Yang et al.~\cite{PreTouchTargetSelYang2011} used pre-touch finger distance to expand nearby targets on screen, facilitating easier target selection. This general approach has not yet been explored in the context of authentication techniques resistant to shoulder surfing.

Another common application of pre-touch information is for reducing the perceived latency of touchscreen interactions. Xia et al. employed this approach for tabletop displays~\cite{Xia0Latency2014}, achieving a touch location prediction error of about 1\,cm. The approach was implemented by tracking the user's index finger location using motion capture with fiducial markers, which are small retro-reflective spheres that can be precisely tracked by IR cameras. In the prototype of our 3D Pattern technique, we also use a motion capture system for finger position tracking.

We anticipate pre-touch sensing to become available on commodity smartphones in the near future. In 2016, Hinckley et al.~\cite{PreTouchHinckley2016} explored how a smartphone with a self-capacitance touchscreen could enable pre-touch information to be sensed, and applied this information in various smartphone applications. We envision that our pre-touch PIN entry techniques will be able to be used on smartphones without additional motion tracking hardware.

\section{Goals and Threat Model}

We assume that the attacker is in close vicinity of the victim during authentication or has a video (e.g., from security camera footage) of the victim authenticating, enabling shoulder-surfing attacks. In regards to smudge attacks, the attacker may have a brief period of physical access to the phone after the victim has authenticated, allowing the attacker time to observe smudges on the screen and potentially deduce the credentials. We consider specialized attacks such as acoustic side-channel attacks~\cite{ChengAcousticAttack2018} to be outside the scope of this work.

With this threat model in mind, we propose the following goals for a pre-touch augmented PIN entry technique:
\begin{itemize}
\item protect against shoulder-surfing and smudge attacks;
\item leverage pre-touch information of the user's finger;
\item not involve significant extra cognitive load or memorization on the part of the user (e.g., unlike previous acoustic PIN entry techniques, which rely on the user remembering sound associations); and
\item be not significantly slower or more error-prone than other conventional techniques (e.g., PIN, Android pattern lock).
\end{itemize}

\section{PIN Entry Techniques}
\label{sec:techniques}

In this section, we describe in detail the three techniques we implemented for PIN entry: PIN, pattern, and 3D Pattern.

Our implementations of the PIN and pattern locks are designed to be as similar as possible to Android's built-in PIN and pattern locks, respectively. For the PIN keypad, we present the numbers zero through nine in a grid along with the masked-out PIN on the screen. For the pattern lock, we present a $3\times3$ grid of points on the screen, which users connect together by dragging their finger from point to point. A line is drawn between the points that the user connects.

\subsection{3D Pattern Lock}
\label{sec:3dlocktechnique}

\begin{figure}[t]
    \centering
    \includegraphics[width=0.8\columnwidth]{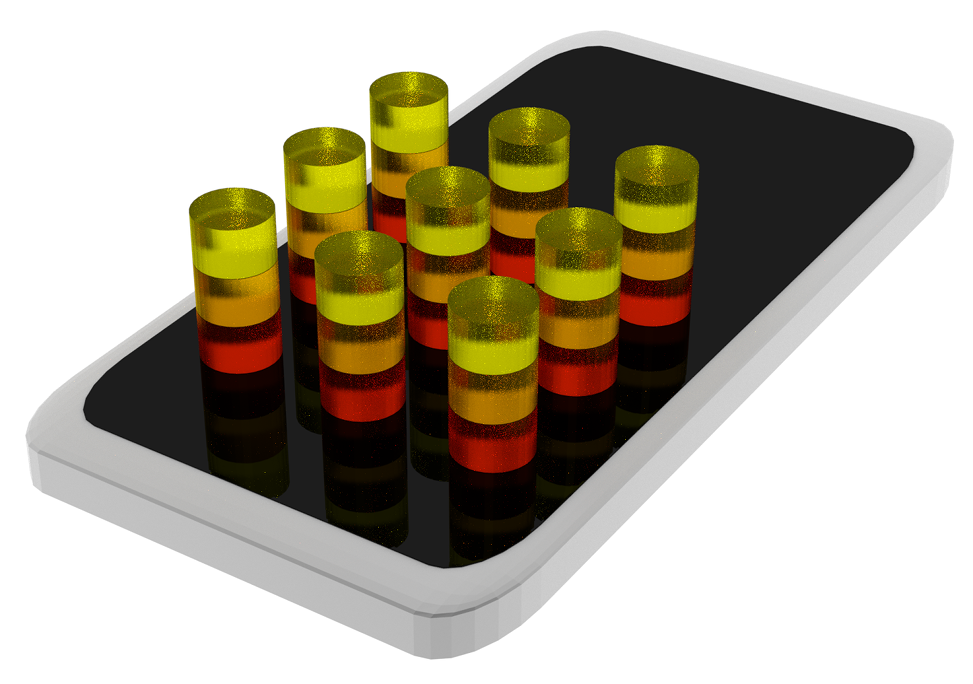}
    \caption{Cylinder representation of our implementation of the 3D Pattern technique. The cylinders are coloured depending on the layer they represent. Users must only hold their finger within the layer of the cylinder to select the corresponding point during authentication.}
    \label{fig:cylinderVersion}
\end{figure}

\begin{figure}[t]
\centering
\subfloat[A screenshot of the smartphone screen as the user sees it before authenticating using the 3D Pattern technique.\label{fig:screenview}]
{{\includegraphics[width=0.45\columnwidth]{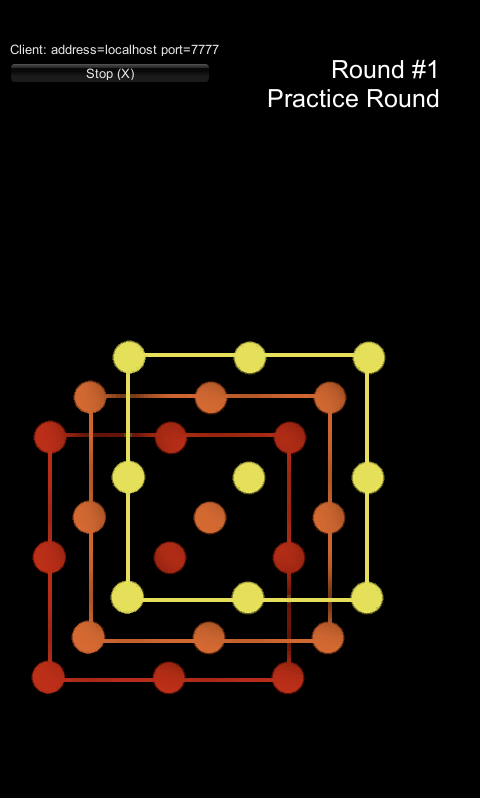} }}~
\subfloat[An example reference image shown to a participant when authenticating using the 3D Pattern technique.\label{fig:referencepat}]
{{\includegraphics[width=0.45\columnwidth]{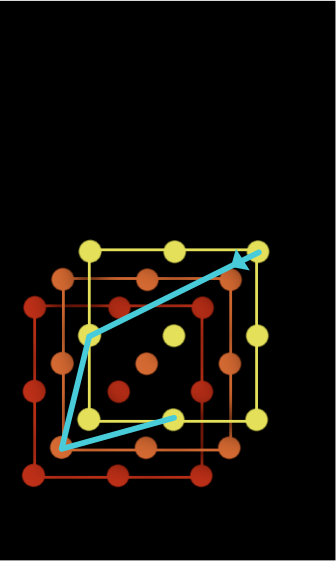} }}\caption{The on-screen representation of the 3D Pattern technique. The 3D cube is orthogonally projected on the screen, with each layer represented using a different colour.}

\end{figure}

The 3D Pattern lock is inspired by Android's conventional pattern lock.
A simple adaptation of the pattern lock to a pre-touch environment would be to duplicate the normal $3\times3$ grid of points, but have users enter the pattern with their finger hovering over the screen rather than touching it, thus eliminating the smudge attack vector. This simple ``hover pattern'' use of pre-touch should maintain most of the security and usability properties of the normal pattern lock, save for being immune to smudge attacks, and so we do not analyze this technique in detail.
Instead, our novel idea is to use pre-touch information to extend the pattern lock concept into a full third ``z'' dimension.
Rather than a $3\times3$ grid of circles, our 3D Pattern lock is a $3\times3\times3$ cube of cylinders (see Figure~\ref{fig:cylinderVersion}). By using cylinders, users do not have to think about where to place their fingers within a layer, only about which layer to place their finger within. This simplifies the rendering of depth cues.
The smartphone renders an orthogonal projection of the cube (see Figure~\ref{fig:screenview}). Each depth, or ``layer'', of the cube is represented in a different colour. The user authenticates by connecting the points in a chosen sequence. Our technique does not require users to slide their fingers on the screen. This inherently protects against smudge attacks since users will not leave oily residues on the screen.

Assuming that the user is allowed to connect any four points such that no point is reused, a theoretical password space upper bound is $27\times26\times24\times23=387504$ patterns. 
However, as a further improvement to usability, we limit the space of valid 3D patterns to include only those that start on the topmost layer, do not bypass the middle layer, do not bypass a point within a layer or use a point more than once (as with the traditional Android pattern lock), and do not connect points across a layer with a distance of more than $\sqrt{3}$ units (to avoid difficult-to-input diagonal lines).
Even with these assumptions, the password space for the 3D Pattern technique is still larger than previous conventional techniques. Using a recursive algorithm in Python, we found 19192 possible 3D Patterns.
This means that the worst-case password space for our 3D Pattern technique is better than that of both PIN (10000) and pattern (1400, computed using a similar Python script as above) locks in the case of a four-digit PIN or pattern.

As with the conventional Android pattern lock, our 3D Pattern lock has two modes: (1)~\emph{with feedback} and (2)~\emph{without feedback}. In \emph{with feedback} mode, as the user moves his or her finger between the different points, a line is rendered between each of the connected points. In \emph{no feedback} mode, these lines are not rendered.

All three of our implemented techniques have additional haptic feedback. This helps the users perceive whether their input has been detected. In a very quiet environment, an attacker may be able to hear the haptic feedback, but in the ambient noise of the experiment room, the experimenters could not hear or otherwise detect the haptic feedback as participants were authenticating. Similarly, an attacker should not be able to hear the haptic feedback in a public environment.

\begin{figure}[p]
\centering
\subfloat[Starting authentication]
{{\includegraphics[width=0.33\textwidth, trim={0 0.2cm 0 0}, clip]{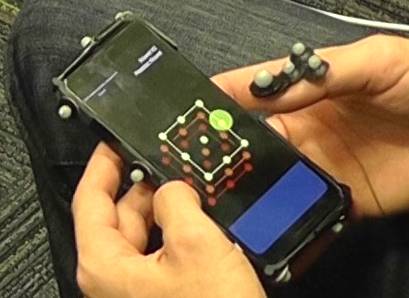} }}

\subfloat[Connecting second point]
{{\includegraphics[width=0.33\textwidth, trim={0 0.2cm 0 0}, clip]{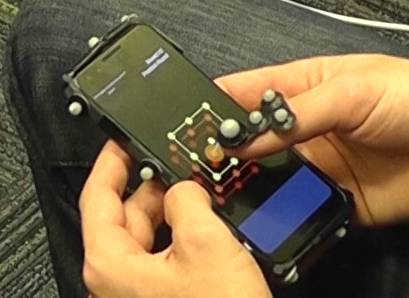} }}

\subfloat[Connecting third point]
{{\includegraphics[width=0.33\textwidth, trim={0 0.2cm 0 0}, clip]{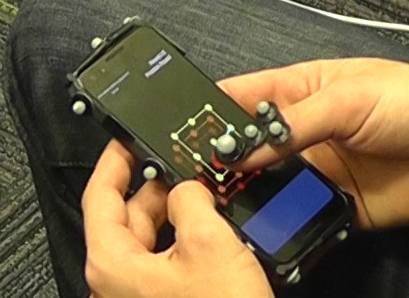} }}

\subfloat[Finished authentication]
{{\includegraphics[width=0.33\textwidth, trim={0 0.2cm 0 0}, clip]{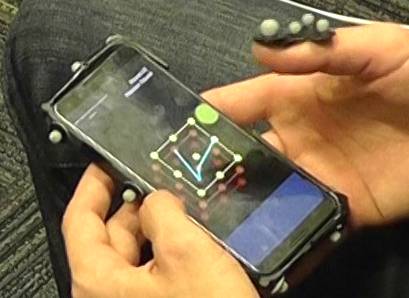} }}

\caption{Study participant authenticating using our 3D Pattern technique. (a) The participant's finger starts in the yellow layer, the closest layer to the participant. (b) The participant moves his finger diagonally to a point in the middle layer. (c) The participant moves his finger to the point below in the bottom layer. (d) The participant moves his finger diagonally up to a point in the middle layer, and the complete pattern has been entered.}
\label{fig:entering_pins}
\end{figure}

Our 3D Pattern lock additionally always renders a ``cursor'' on the screen. The cursor changes colour depending on the finger's distance from the screen. Yellow represents that the finger is in the closest layer to the user, orange represents the middle layer, and red represents the layer closest to the screen. Figure~\ref{fig:entering_pins} demonstrates the authentication process.

\section{Implementation}
\label{sec:impl}

\begin{figure*}[t]
    \centering
    \includegraphics[width=0.8\textwidth]{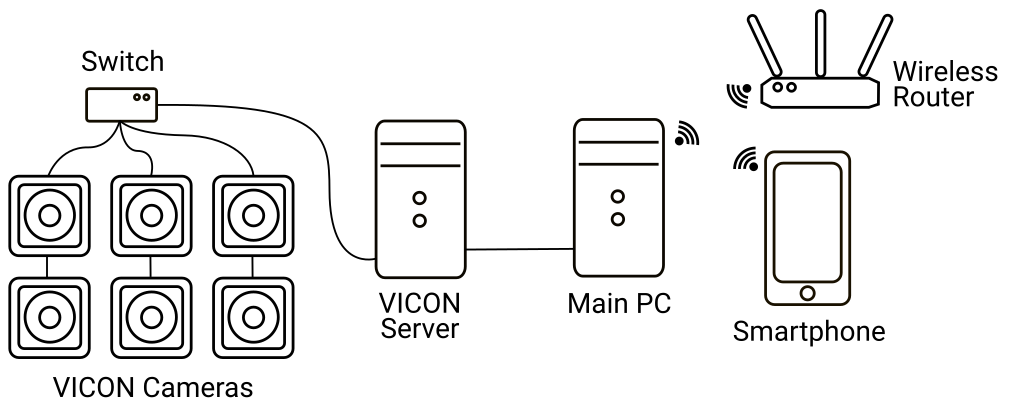}
    \caption{System architecture diagram of the prototype smartphone authentication techniques. Information flow is generally from the left to the right. The Vicon cameras capture the position of fiducial markers in 3D space, and forward this information to a Vicon server through a network switch. The Vicon server calculates the absolute positions of the smartphone and the user's finger, and forwards this over a link-local connection to the main PC. The main PC establishes a bidirectional Wi-Fi connection with the smartphone through a router. It uses input data from the smartphone in conjunction with motion capture object positions to operate the techniques. In practice, when pre-touch technology becomes commercially available, all computation and sensing will be performed on the smartphone itself.}
    \label{fig:diagram}
\end{figure*}

Pre-touch information is not yet available on current commercial smartphones. As a result, we simulated pre-touch with fiducial-based motion capture. Our system was set up in a $2\,\mathrm m\times2\,\mathrm m\times3\,\mathrm m$ room instrumented with six Vicon motion-capture cameras. The cameras work together to triangulate the position of fiducial markers on the user's finger. The small grey spheres in Figure~\ref{fig:entering_pins} are examples of these markers. The motion capture system tracks both the smartphone and the finger positions. The absolute positions of each of these objects in 3D space is transformed, resulting in finger coordinates relative to the phone screen.

Our prototype implementation includes a main PC, which instruments the remaining parts of the system. The responsibilities of the main PC include determining what to draw on the phone screen, controlling the experiment, verifying PINs, and logging useful information. The three authentication techniques were implemented using the Unity game engine.\footnote{\url{https://unity3d.com}}
The source code is available at
\url{https://github.com/spamalot/3D-Pattern-Lock}.
All six Vicon cameras are connected to a Vicon server through a network switch. This server calculates the absolute 3D positions of the user's finger and smartphone and forwards this information to the main PC. The connection between the smartphone and the main PC is implemented as a client-server architecture over a Wi-Fi connection. The smartphone renders the authentication technique to the user and accepts touch and dragging input. The system architecture is depicted in Figure~\ref{fig:diagram}. If our system were to be implemented in practice, of course, all computation and sensing would be performed on the smartphone itself.

To evaluate our technique, we compare it to two other popular authentication techniques: PIN and pattern locks. We replicate Android's implementation of the techniques as closely as possible to make a fair evaluation of our proposed technique. We do this by rendering Android's layout as a background, and overlaying buttons or swipe zones on top of this image. This preserves the spacing between UI elements present in the Android implementations. For the Pattern lock, we empirically matched the width of each point's hit box with that of the Android implementation.

\section{Experiment}

\begin{figure}[t]
    \centering
    \includegraphics[width=\columnwidth]{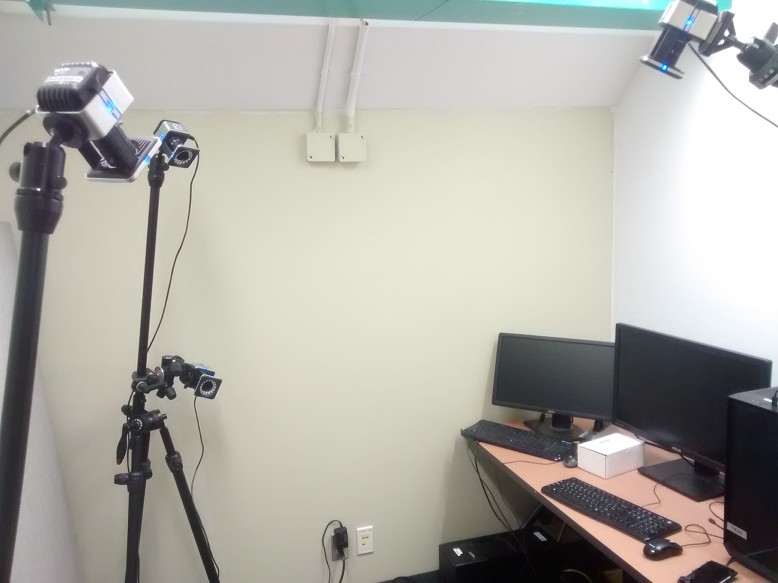}
    \caption{Room setup used for the experiment. Six Vicon cameras were placed throughout the room. Participants sat on a chair against the far wall, with a video camera pointing over their left shoulder.}
    \label{fig:experiment}
\end{figure}

We conducted an experiment to understand how quickly and accurately users can authenticate using the 3D Pattern technique, as well as the resistance of our technique against shoulder surfing. Our hypotheses are that the 3D Pattern technique will be slightly slower and less accurate than the PIN and pattern techniques, but significantly more shoulder surfing resistant. We received approval from our university's research ethics board for this experiment involving human participants
(ORE\# 22114).

We recruited six graduate student participants from the university community via word of mouth. Participants each received a gift card valued at \$5 for their participation in the 30-minute session.

\subsection{Participants}

We recruited 6 participants (5 male, 1 female) with average age 24 (SD=2) to enter PINs on a mobile phone. All participants were right-handed and used their right hand to authenticate using the three implemented techniques (PIN, pattern, and 3D Pattern). Participants reported using a diverse range of locks on their personal phones, with one participant using a PIN keypad, another using Android pattern lock, three using fingerprint reading, and one using iris scanning.

\subsection{Apparatus}

The previously described implementation was used for the experiment. Figure~\ref{fig:experiment} shows the general layout of the experiment room. Participants sat against a wall with no motion capture cameras, to minimize occlusion of the cameras. Participants authenticated on a Google Pixel 3 smartphone. A 1080p video camera was mounted on a tripod, aiming over the left shoulder
of the participant. This camera provided a clear view of the phone screen and was used to mimic the view of a shoulder surfer.

\subsection{Task}

The experiment was divided into two sections. In the first section, the participant was instructed to authenticate using each of the three techniques. Input events on the smartphone were logged on the main PC and videos of participants authenticating were recorded.

All authentication techniques were four ``digits'' long; that is, PINs had four numbers and patterns involved connecting four points. Each point in the pattern corresponded to a digit. For 3D Pattern lock, the top-left point of each layer corresponded to the digits 0, 9, and 18. Within the same layer, the digits increased left to right, top to bottom. For example, in the layer closest to the screen, the top-left point corresponded to 0, the top-middle point corresponded to 1, and so on.

For the second section, the participant was asked to shoulder surf the videos of the previous participant authenticating; the last successful (most practiced) authentication of each PIN or pattern for each technique was shown. The first participant performed shoulder surfing once the last participant finished authenticating using all techniques. The participant had up to 20 guesses to correctly determine the PIN or pattern entered. While guessing, participants were allowed to consult reference images of each authentication technique (e.g., see Figure~\ref{fig:screenview}).
We chose to have our participants, who used our 3D Pattern technique, be the shoulder surfers because they were familiar with this novel technique.

\subsection{Design and Procedure}

The study was a within-subjects design with
\textsc{technique} and \textsc{trial number} as independent variables. Technique \textit{Entry Time}, technique \textit{Error Rate}, and shoulder surfing \textit{Guesses} were measured as dependent variables. \textsc{Technique} had 3 levels: PIN, Pattern, and Pattern3D, the last of which corresponded to our 3D Pattern design.

Techniques were presented to participants in a Latin square arrangement. For each technique, there were two PINs or patterns. For each PIN or pattern, there were two blocks of authentication trials, each with five trials. The first block of each PIN or pattern was a practice round, and the data was not analyzed. The practice round of Pattern and Pattern3D rendered in \emph{with feedback} mode, which rendered lines between connected points on the screen. However, during the second block, the technique rendered in \emph{without feedback} mode.

PINs and patterns were randomly generated. During the first section of the experiment, participants were allowed to look at a separate computer monitor on which the reference PINs and patterns were displayed. The pattern and 3D Pattern reference images were rendered as they would be seen after being entered on the phone screen (see Figure~\ref{fig:referencepat}).
As described in Section~\ref{sec:3dlocktechnique}, to improve usability, patterns for the Pattern3D technique were controlled to always start on the top (furthest from screen) layer.

\section{Results}

\begin{figure}[t]
    \centering
    \includegraphics[width=\columnwidth]{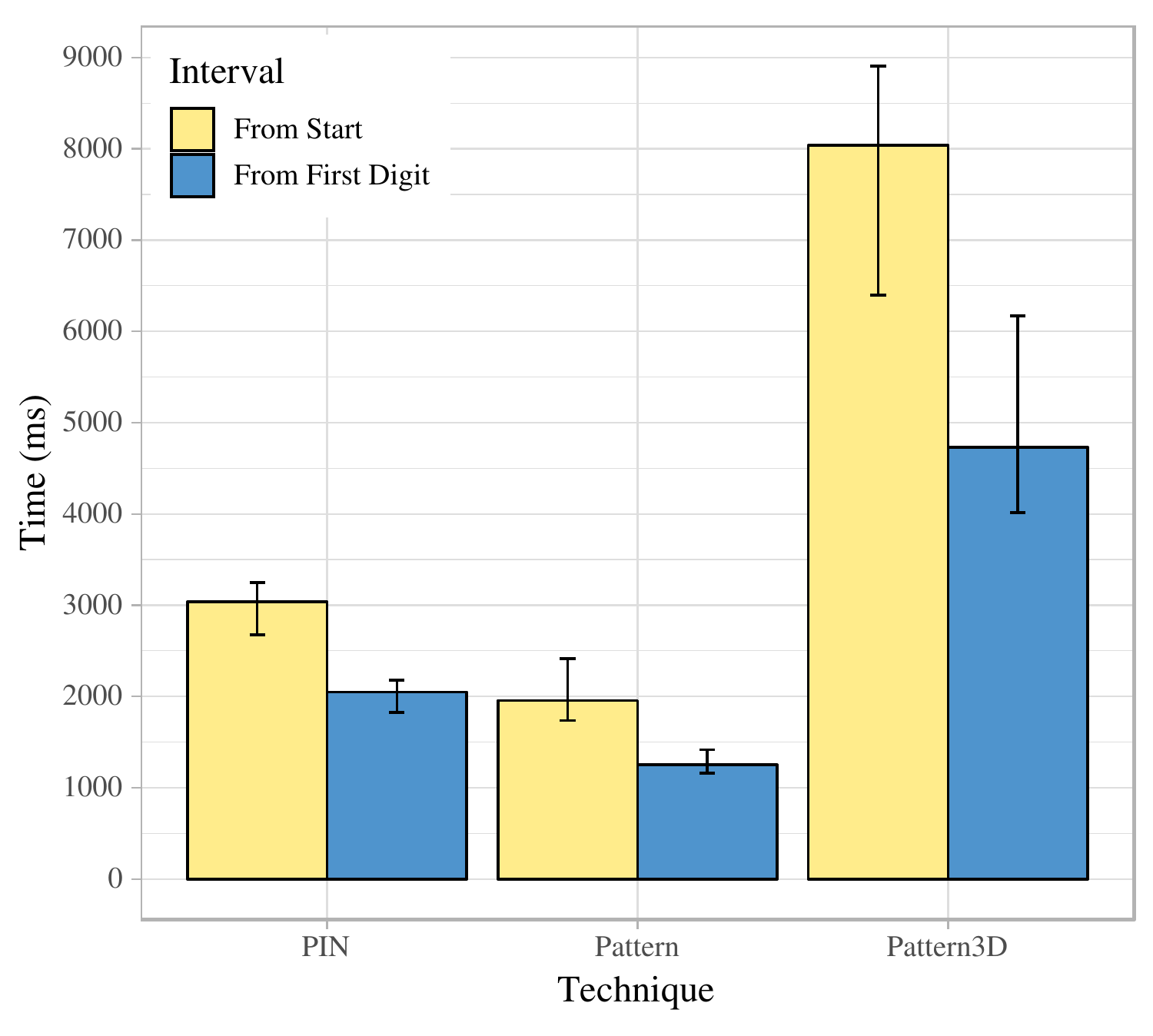}
    \caption{Median times taken to authenticate for each of the authentication techniques. ``From Start'' indicates the time from the user pressing the start button to finishing authentication; ``From First Digit'' indicates the time from the user entering the first digit to finishing authentication. Error bars indicate 95\% confidence intervals.}
    \label{fig:time}
\end{figure}

A repeated measures ANOVA with Greenhouse-Geisser sphericity correction found a significant main effect of \textsc{technique} on log-transformed \textit{Entry Time} \anova{1.36}{6.84}{22.79}{0.01}{0.71}. Post hoc paired t-tests with Holm correction show with significance that Pattern3D was slower than PIN \p{0.0001} and Pattern \p{0.0001}, and that PIN was slower than Pattern \p{0.001}. The median time to authenticate using PIN was 3.0 seconds (IQR=1.2), Pattern was 2.0 seconds (IQR=1.5), and Pattern3D was 8.0 seconds (IQR=7.1). We also measured the ``time from first digit'', or the difference in time between the first input towards authenticating and finishing authentication. The median time from first digit using PIN was 2.0 seconds (IQR=1.0), Pattern was 1.3 seconds (IQR=0.8), and Pattern3D was 4.7 seconds (IQR=4.4). These results are depicted in Figure~\ref{fig:time}.

\begin{figure}[t]
    \centering
    \includegraphics[width=\columnwidth]{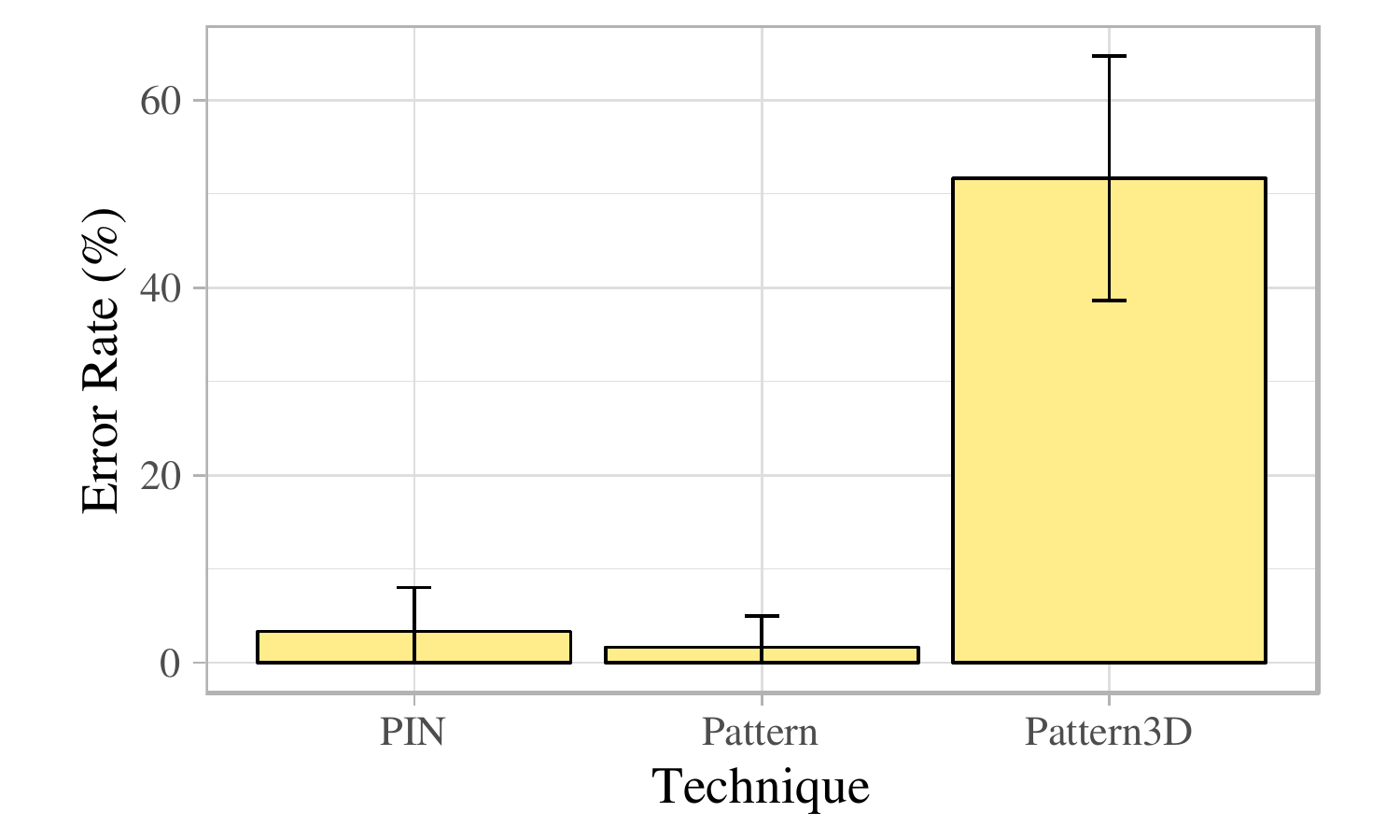}
    \caption{Mean error rates for each of the authentication techniques. Error bars indicate 95\% confidence intervals.}
    \label{fig:errRate}
\end{figure}

A Friedman rank sum test shows a significant effect of \textsc{technique} on \textit{Error Rate} ($\chi^2_3 = 17.72, p < 0.001$). Post hoc paired Wilcoxon signed-rank tests with Holm correction show with significance that Pattern3D has a higher error rate than PIN \p{0.0001} and Pattern \p{0.0001}, but do not indicate any significant difference between PIN and Pattern \peq{0.57}. The mean error rate for PIN was 3\% (SD=18\%), Pattern was 2\% (SD=13\%), and Pattern3D was 52\% (SD=50\%). These results are depicted in Figure~\ref{fig:errRate}.

\begin{figure}[t]
    \centering
    \includegraphics[width=\columnwidth]{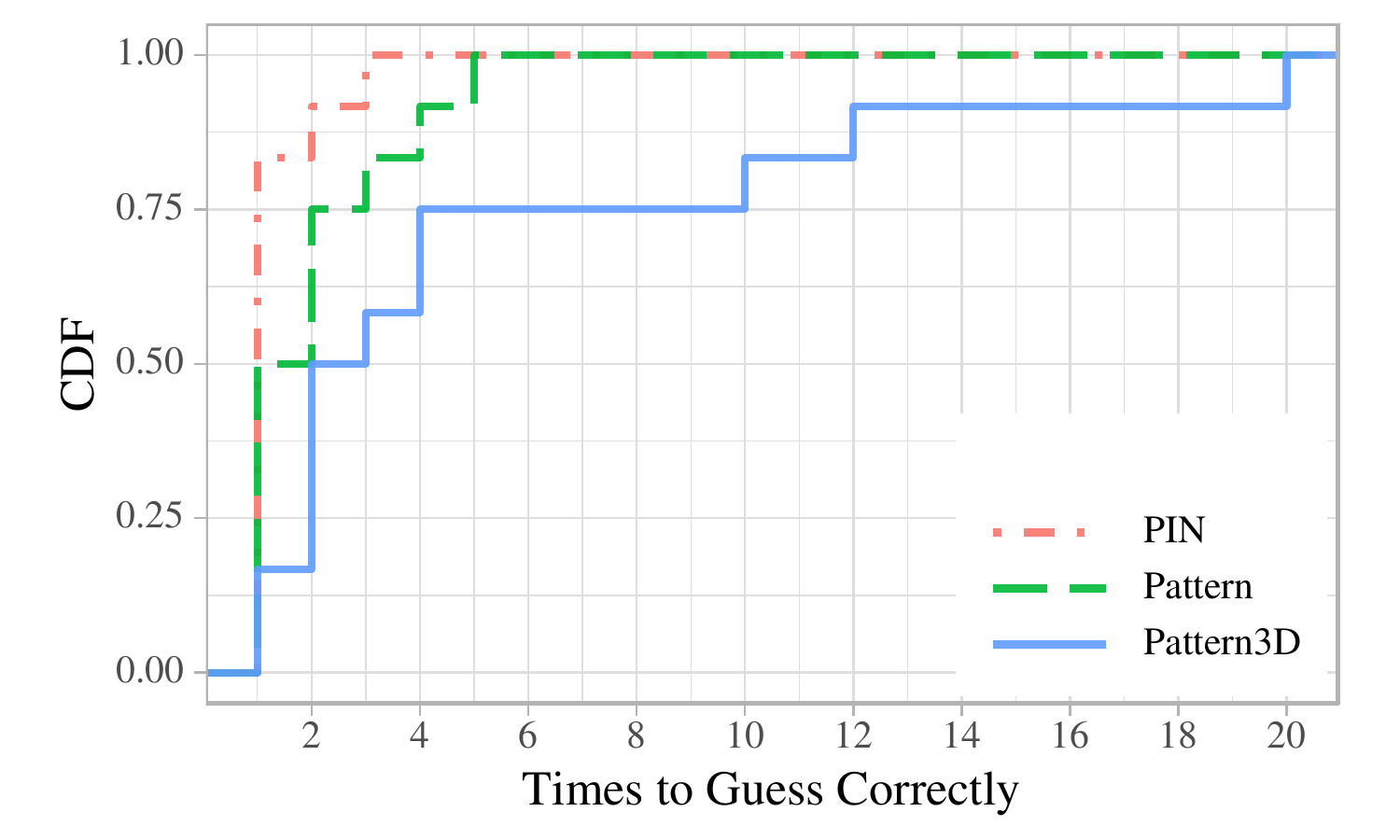}
    \caption{Empirical CDF of number of guesses needed to correctly identify the PIN or pattern in the shoulder surfing video.}
    \label{fig:cdf}
\end{figure}

An empirical CDF representing the number of guesses needed to correctly guess a PIN or pattern from a shoulder-surfing video is depicted in Figure~\ref{fig:cdf}.
PIN and Pattern are similar in shoulder-surfing resistance, whereas Pattern3D appears to have a slight advantage. A Friedman rank sum test shows a significant effect of \textsc{technique} on \textit{Guesses} ($\chi^2_3 = 9.4, p < 0.05$). Post hoc paired Wilcoxon signed-rank tests with Holm correction show with significance that Pattern3D was harder to guess correctly than PIN \p{0.01}, but shows no other significant effects. The mean number of guesses needed for PIN was 1.3 (SD=1, median=1), Pattern was 2.0 (SD=1, median=1.5), and Pattern3D was 5.3 (SD=6, median=2.5). One participant was not able to guess one 3D Pattern within the given 20 trials.

\section{Discussion}

Based on our experience designing and evaluating the 3D Pattern technique, we discuss the shoulder-surfing resistance of the technique, and possible future directions for exploration.

\subsection{Shoulder-Surfing Resistance}

Statistical analysis shows that the shoulder-surfing resistance of the 3D Pattern technique is higher than that of PIN locks. The empirical CDF of shoulder-surfing guesses (Figure~\ref{fig:cdf}) might indicate that the 3D Pattern technique is more shoulder surfing resistant than the pattern technique. Because users hover their fingers in 3D space, it is hard for a shoulder surfer to guess the layer in which the user's finger hovers. Moreover, due to the fact that the pattern and 3D Pattern locks rendered in \emph{without feedback} mode, no lines were rendered on the screen. This could explain why pattern lock appeared slightly more shoulder surfing resistant than the PIN technique.

Previous work~\cite{OladePictures2018,TakadaAwaseE2003} has used graphical visualisation to provide more memorable passwords. Our technique could be expanded to combine the ideas of both picture passwords and pre-touch information. This would allow for the creation of meaningful and easily remembered passwords that are less position sensitive.

\subsection{Equipment Resolution and Latency} The Vicon tracking system has the potential to be very accurate when tracking, but this depends on a number of environmental factors such as lighting, camera placement, and so on. In the study at hand, there was some visible jitter in the position of the tracked objects, likely slowing down participants when using the 3D Pattern technique. While not reported by participants, wearing the fiducial marker with double-sided tape on the finger may have caused discomfort and also slowed down authentication time. Further, due to network latency and the high frequency of messages being sent across the network, user interface latency was low
but apparent with all three techniques, potentially affecting the external validity of our study.
In the future, smartphones with built-in pre-touch support would eliminate the need for motion capture and its associated limitations.

\subsection{Visualization of the Cube} Representing the 3D cube on the 2D phone screen is important for our technique to be effective. Our choice to use orthogonal projection may have negatively affected our results both in terms of authentication speed and error rate.

One possible way to extend our design would be to use different kinds of depth cues. Our implementation uses a cursor with varying colours and sizes based on depth. We could instead visualize shadows to give an impression of finger height. The distance between the cursor and its shadow and shadow size could be varied with finger height. Alternatively, a depth-of-field blurring effect based on the finger's distance from the screen could be used. Another possible extension would be to investigate if the use of perspective projection instead of our implemented orthogonal projection would have better performance. It could also be effective to slightly rotate the projection of the cube depending on the position of the finger or orientation of the smartphone.

\subsection{Experimental Protocol}

Our experiment allowed the user to practice each technique five times. Given the novelty of pre-touch interfaces, this might not have been enough practice rounds for users to become accustomed to this new paradigm. Further in support of this argument is the fact that participants took several seconds between starting authentication and entering the first digit for the 3D Pattern technique. Users needed to adjust the position of their fingers to find the correct position of the top layer.

Compared to previous studies~\cite{HarbachAnatomy2016}, both our implementations of PIN authentication and pattern authentication were slower. The are several factors that could have contributed to these results. First, the main PC, rather than the smartphone, performed all calculations, resulting in a possible small effect of network latency. We also found that participants sometimes referred back to the secondary computer monitor to recall which PIN or pattern to enter.

\section{Conclusion}

In this work, we have proposed a novel approach to smartphone authentication using pre-touch information, called 3D Pattern. We have implemented a prototype of the technique by simulating a pre-touch-capable smartphone using a motion capture system. We have also evaluated the 3D Pattern technique in a pilot study, in comparison to two popular existing techniques, finding that authentication times were longer, but that the technique could be more resistant to shoulder-surfing attacks, while being immune to smudge attacks. We attribute the longer authentication times to environmental conditions adversely affecting motion capture and the novelty of pre-touch to participants. We believe these limitations could be easily overcome as pre-touch becomes mainstream.

\section*{Acknowledgements}

This work was made possible by NSERC Discovery Grants 2016-03878,
2017-03858, and 2018-05187, the Canada Foundation for Innovation
Infrastructure Fund ``Facility for Fully Interactive Physio-digital
Spaces'' (\#33151), and Ontario Early Researcher Award \#ER16-12-184.

\bibliographystyle{ACM-Reference-Format}
\bibliography{references}


\begin{thebibliography}{27}


\ifx \showCODEN    \undefined \def \showCODEN     #1{\unskip}     \fi
\ifx \showDOI      \undefined \def \showDOI       #1{#1}\fi
\ifx \showISBNx    \undefined \def \showISBNx     #1{\unskip}     \fi
\ifx \showISBNxiii \undefined \def \showISBNxiii  #1{\unskip}     \fi
\ifx \showISSN     \undefined \def \showISSN      #1{\unskip}     \fi
\ifx \showLCCN     \undefined \def \showLCCN      #1{\unskip}     \fi
\ifx \shownote     \undefined \def \shownote      #1{#1}          \fi
\ifx \showarticletitle \undefined \def \showarticletitle #1{#1}   \fi
\ifx \showURL      \undefined \def \showURL       {\relax}        \fi
\providecommand\bibfield[2]{#2}
\providecommand\bibinfo[2]{#2}
\providecommand\natexlab[1]{#1}
\providecommand\showeprint[2][]{arXiv:#2}

\bibitem[\protect\citeauthoryear{Aviv, Gibson, Mossop, Blaze, and Smith}{Aviv
  et~al\mbox{.}}{2010}]%
        {AvivSmudgeAttack2010}
\bibfield{author}{\bibinfo{person}{Adam~J. Aviv}, \bibinfo{person}{Katherine
  Gibson}, \bibinfo{person}{Evan Mossop}, \bibinfo{person}{Matt Blaze}, {and}
  \bibinfo{person}{Jonathan~M. Smith}.} \bibinfo{year}{2010}\natexlab{}.
\newblock \showarticletitle{Smudge Attacks on Smartphone Touch Screens}. In
  \bibinfo{booktitle}{\emph{Proceedings of the 4th USENIX Conference on
  Offensive Technologies}} \emph{(\bibinfo{series}{WOOT'10})}.
  \bibinfo{publisher}{USENIX Association}, \bibinfo{address}{Berkeley, CA,
  USA}, \bibinfo{pages}{1--7}.
\newblock
\urldef\tempurl%
\url{http://dl.acm.org/citation.cfm?id=1925004.1925009}
\showURL{%
\tempurl}


\bibitem[\protect\citeauthoryear{Bianchi, Oakley, Kostakos, and Kwon}{Bianchi
  et~al\mbox{.}}{2011}]%
        {BianchiHaptic2011}
\bibfield{author}{\bibinfo{person}{Andrea Bianchi}, \bibinfo{person}{Ian
  Oakley}, \bibinfo{person}{Vassilis Kostakos}, {and} \bibinfo{person}{Dong~Soo
  Kwon}.} \bibinfo{year}{2011}\natexlab{}.
\newblock \showarticletitle{The Phone Lock: Audio and Haptic Shoulder-surfing
  Resistant PIN Entry Methods for Mobile Devices}. In
  \bibinfo{booktitle}{\emph{Proceedings of the Fifth International Conference
  on Tangible, Embedded, and Embodied Interaction}} \emph{(\bibinfo{series}{TEI
  '11})}. \bibinfo{publisher}{ACM}, \bibinfo{address}{New York, NY, USA},
  \bibinfo{pages}{197--200}.
\newblock
\showISBNx{978-1-4503-0478-8}
\urldef\tempurl%
\url{https://doi.org/10.1145/1935701.1935740}
\showDOI{\tempurl}


\bibitem[\protect\citeauthoryear{Buschek, De~Luca, and Alt}{Buschek
  et~al\mbox{.}}{2015}]%
        {UserPatternsTypingPasswordBuschek2015}
\bibfield{author}{\bibinfo{person}{Daniel Buschek}, \bibinfo{person}{Alexander
  De~Luca}, {and} \bibinfo{person}{Florian Alt}.}
  \bibinfo{year}{2015}\natexlab{}.
\newblock \showarticletitle{Improving Accuracy, Applicability and Usability of
  Keystroke Biometrics on Mobile Touchscreen Devices}. In
  \bibinfo{booktitle}{\emph{Proceedings of the 33rd Annual ACM Conference on
  Human Factors in Computing Systems}} \emph{(\bibinfo{series}{CHI '15})}.
  \bibinfo{publisher}{ACM}, \bibinfo{address}{New York, NY, USA},
  \bibinfo{pages}{1393--1402}.
\newblock
\showISBNx{978-1-4503-3145-6}
\urldef\tempurl%
\url{https://doi.org/10.1145/2702123.2702252}
\showDOI{\tempurl}


\bibitem[\protect\citeauthoryear{Cheng, Bagci, Roedig, and Yan}{Cheng
  et~al\mbox{.}}{2018}]%
        {ChengAcousticAttack2018}
\bibfield{author}{\bibinfo{person}{Peng Cheng}, \bibinfo{person}{Ibrahim
  Bagci}, \bibinfo{person}{Utz Roedig}, {and} \bibinfo{person}{Jeff Yan}.}
  \bibinfo{year}{2018}\natexlab{}.
\newblock \bibinfo{title}{SonarSnoop: Active Acoustic Side-Channel Attacks}.
\newblock \bibinfo{howpublished}{https://arxiv.org/abs/1808.10250}.
\newblock
\urldef\tempurl%
\url{https://arxiv.org/abs/1808.10250}
\showURL{%
\tempurl}


\bibitem[\protect\citeauthoryear{Chiang and Chiasson}{Chiang and
  Chiasson}{2013}]%
        {GraphicalPassowordsChiang2013}
\bibfield{author}{\bibinfo{person}{Hsin-Yi Chiang} {and} \bibinfo{person}{Sonia
  Chiasson}.} \bibinfo{year}{2013}\natexlab{}.
\newblock \showarticletitle{Improving User Authentication on Mobile Devices: A
  Touchscreen Graphical Password}. In \bibinfo{booktitle}{\emph{Proceedings of
  the 15th International Conference on Human-computer Interaction with Mobile
  Devices and Services}} \emph{(\bibinfo{series}{MobileHCI '13})}.
  \bibinfo{publisher}{ACM}, \bibinfo{address}{New York, NY, USA},
  \bibinfo{pages}{251--260}.
\newblock
\showISBNx{978-1-4503-2273-7}
\urldef\tempurl%
\url{https://doi.org/10.1145/2493190.2493213}
\showDOI{\tempurl}


\bibitem[\protect\citeauthoryear{De~Luca, Weiss, and Hussmann}{De~Luca
  et~al\mbox{.}}{2007}]%
        {DeLucaPatternLock2007}
\bibfield{author}{\bibinfo{person}{Alexander De~Luca}, \bibinfo{person}{Roman
  Weiss}, {and} \bibinfo{person}{Heinrich Hussmann}.}
  \bibinfo{year}{2007}\natexlab{}.
\newblock \showarticletitle{PassShape: Stroke Based Shape Passwords}. In
  \bibinfo{booktitle}{\emph{Proceedings of the 19th Australasian Conference on
  Computer-Human Interaction: Entertaining User Interfaces}}
  \emph{(\bibinfo{series}{OZCHI '07})}. \bibinfo{publisher}{ACM},
  \bibinfo{address}{New York, NY, USA}, \bibinfo{pages}{239--240}.
\newblock
\showISBNx{978-1-59593-872-5}
\urldef\tempurl%
\url{https://doi.org/10.1145/1324892.1324943}
\showDOI{\tempurl}


\bibitem[\protect\citeauthoryear{Eiband, Khamis, von Zezschwitz, Hussmann, and
  Alt}{Eiband et~al\mbox{.}}{2017}]%
        {EibandSurfingWild2017}
\bibfield{author}{\bibinfo{person}{Malin Eiband}, \bibinfo{person}{Mohamed
  Khamis}, \bibinfo{person}{Emanuel von Zezschwitz}, \bibinfo{person}{Heinrich
  Hussmann}, {and} \bibinfo{person}{Florian Alt}.}
  \bibinfo{year}{2017}\natexlab{}.
\newblock \showarticletitle{Understanding Shoulder Surfing in the Wild: Stories
  from Users and Observers}. In \bibinfo{booktitle}{\emph{Proceedings of the
  2017 CHI Conference on Human Factors in Computing Systems}}
  \emph{(\bibinfo{series}{CHI '17})}. \bibinfo{publisher}{ACM},
  \bibinfo{address}{New York, NY, USA}, \bibinfo{pages}{4254--4265}.
\newblock
\showISBNx{978-1-4503-4655-9}
\urldef\tempurl%
\url{https://doi.org/10.1145/3025453.3025636}
\showDOI{\tempurl}


\bibitem[\protect\citeauthoryear{Frank, Biedert, Ma, Martinovic, and
  Song}{Frank et~al\mbox{.}}{2013}]%
        {FrankTouchalytics2013}
\bibfield{author}{\bibinfo{person}{Mario Frank}, \bibinfo{person}{Ralf
  Biedert}, \bibinfo{person}{Eugene Ma}, \bibinfo{person}{Ivan Martinovic},
  {and} \bibinfo{person}{Dawn Song}.} \bibinfo{year}{2013}\natexlab{}.
\newblock \showarticletitle{Touchalytics: On the Applicability of Touchscreen
  Input as a Behavioral Biometric for Continuous Authentication}.
\newblock \bibinfo{journal}{\emph{IEEE Transactions on Information Forensics
  and Security}} \bibinfo{volume}{8}, \bibinfo{number}{1}
  (\bibinfo{date}{January} \bibinfo{year}{2013}), \bibinfo{pages}{136--148}.
\newblock
\showISSN{1556-6013}
\urldef\tempurl%
\url{https://doi.org/10.1109/TIFS.2012.2225048}
\showDOI{\tempurl}


\bibitem[\protect\citeauthoryear{Guerar, Merlo, and Migliardi}{Guerar
  et~al\mbox{.}}{2017}]%
        {GuerarFakePattern2017}
\bibfield{author}{\bibinfo{person}{Meriem Guerar}, \bibinfo{person}{Alessio
  Merlo}, {and} \bibinfo{person}{Mauro Migliardi}.}
  \bibinfo{year}{2017}\natexlab{}.
\newblock \showarticletitle{Clickpattern: A pattern lock system resilient to
  smudge and side-channel attacks}.
\newblock \bibinfo{journal}{\emph{Journal of Wireless Mobile Networks,
  Ubiquitous Computing, and Dependable Applications}}  \bibinfo{volume}{8}
  (\bibinfo{date}{January} \bibinfo{year}{2017}), \bibinfo{pages}{64--78}.
\newblock


\bibitem[\protect\citeauthoryear{Harbach, De~Luca, and Egelman}{Harbach
  et~al\mbox{.}}{2016}]%
        {HarbachAnatomy2016}
\bibfield{author}{\bibinfo{person}{Marian Harbach}, \bibinfo{person}{Alexander
  De~Luca}, {and} \bibinfo{person}{Serge Egelman}.}
  \bibinfo{year}{2016}\natexlab{}.
\newblock \showarticletitle{The Anatomy of Smartphone Unlocking: A Field Study
  of Android Lock Screens}. In \bibinfo{booktitle}{\emph{Proceedings of the
  2016 CHI Conference on Human Factors in Computing Systems}}
  \emph{(\bibinfo{series}{CHI '16})}. \bibinfo{publisher}{ACM},
  \bibinfo{address}{New York, NY, USA}, \bibinfo{pages}{4806--4817}.
\newblock
\showISBNx{978-1-4503-3362-7}
\urldef\tempurl%
\url{https://doi.org/10.1145/2858036.2858267}
\showDOI{\tempurl}


\bibitem[\protect\citeauthoryear{Harbach, von Zezschwitz, Fichtner, Luca, and
  Smith}{Harbach et~al\mbox{.}}{2014}]%
        {HarbachPerception2014}
\bibfield{author}{\bibinfo{person}{Marian Harbach}, \bibinfo{person}{Emanuel
  von Zezschwitz}, \bibinfo{person}{Andreas Fichtner},
  \bibinfo{person}{Alexander~De Luca}, {and} \bibinfo{person}{Matthew Smith}.}
  \bibinfo{year}{2014}\natexlab{}.
\newblock \showarticletitle{It{\textquoteright}s a Hard Lock Life: A Field
  Study of Smartphone (Un)Locking Behavior and Risk Perception}. In
  \bibinfo{booktitle}{\emph{10th Symposium On Usable Privacy and Security
  ({SOUPS} 2014)}}. \bibinfo{publisher}{{USENIX} Association},
  \bibinfo{address}{Menlo Park, CA}, \bibinfo{pages}{213--230}.
\newblock
\showISBNx{978-1-931971-13-3}
\urldef\tempurl%
\url{https://www.usenix.org/conference/soups2014/proceedings/presentation/harbach}
\showURL{%
\tempurl}


\bibitem[\protect\citeauthoryear{Hinckley, Heo, Pahud, Holz, Benko, Sellen,
  Banks, O'Hara, Smyth, and Buxton}{Hinckley et~al\mbox{.}}{2016}]%
        {PreTouchHinckley2016}
\bibfield{author}{\bibinfo{person}{Ken Hinckley}, \bibinfo{person}{Seongkook
  Heo}, \bibinfo{person}{Michel Pahud}, \bibinfo{person}{Christian Holz},
  \bibinfo{person}{Hrvoje Benko}, \bibinfo{person}{Abigail Sellen},
  \bibinfo{person}{Richard Banks}, \bibinfo{person}{Kenton O'Hara},
  \bibinfo{person}{Gavin Smyth}, {and} \bibinfo{person}{William Buxton}.}
  \bibinfo{year}{2016}\natexlab{}.
\newblock \showarticletitle{Pre-Touch Sensing for Mobile Interaction}. In
  \bibinfo{booktitle}{\emph{Proceedings of the 2016 CHI Conference on Human
  Factors in Computing Systems}} \emph{(\bibinfo{series}{CHI '16})}.
  \bibinfo{publisher}{ACM}, \bibinfo{address}{New York, NY, USA},
  \bibinfo{pages}{2869--2881}.
\newblock
\showISBNx{978-1-4503-3362-7}
\urldef\tempurl%
\url{https://doi.org/10.1145/2858036.2858095}
\showDOI{\tempurl}


\bibitem[\protect\citeauthoryear{Khan, Hengartner, and Vogel}{Khan
  et~al\mbox{.}}{2016}]%
        {KhanTargetedMimicry2016}
\bibfield{author}{\bibinfo{person}{Hassan Khan}, \bibinfo{person}{Urs
  Hengartner}, {and} \bibinfo{person}{Daniel Vogel}.}
  \bibinfo{year}{2016}\natexlab{}.
\newblock \showarticletitle{Targeted Mimicry Attacks on Touch Input Based
  Implicit Authentication Schemes}. In \bibinfo{booktitle}{\emph{Proceedings of
  the 14th Annual International Conference on Mobile Systems, Applications, and
  Services}} \emph{(\bibinfo{series}{MobiSys '16})}. \bibinfo{publisher}{ACM},
  \bibinfo{address}{New York, NY, USA}, \bibinfo{pages}{387--398}.
\newblock
\showISBNx{978-1-4503-4269-8}
\urldef\tempurl%
\url{https://doi.org/10.1145/2906388.2906404}
\showDOI{\tempurl}


\bibitem[\protect\citeauthoryear{Khan, Hengartner, and Vogel}{Khan
  et~al\mbox{.}}{2018}]%
        {KhanForcePIN2018}
\bibfield{author}{\bibinfo{person}{Hassan Khan}, \bibinfo{person}{Urs
  Hengartner}, {and} \bibinfo{person}{Daniel Vogel}.}
  \bibinfo{year}{2018}\natexlab{}.
\newblock \showarticletitle{Evaluating Attack and Defense Strategies for
  Smartphone PIN Shoulder Surfing}. In \bibinfo{booktitle}{\emph{Proceedings of
  the 2018 CHI Conference on Human Factors in Computing Systems}}
  \emph{(\bibinfo{series}{CHI '18})}. \bibinfo{publisher}{ACM},
  \bibinfo{address}{New York, NY, USA}, Article \bibinfo{articleno}{164},
  \bibinfo{numpages}{10}~pages.
\newblock
\showISBNx{978-1-4503-5620-6}
\urldef\tempurl%
\url{https://doi.org/10.1145/3173574.3173738}
\showDOI{\tempurl}


\bibitem[\protect\citeauthoryear{Kim, Park, Yoon, and Lee}{Kim
  et~al\mbox{.}}{2018}]%
        {FingerIDKim2018}
\bibfield{author}{\bibinfo{person}{Insu Kim}, \bibinfo{person}{Keunwoo Park},
  \bibinfo{person}{Youngwoo Yoon}, {and} \bibinfo{person}{Geehyuk Lee}.}
  \bibinfo{year}{2018}\natexlab{}.
\newblock \showarticletitle{Touch180: Finger Identification on Mobile
  Touchscreen Using Fisheye Camera and Convolutional Neural Network}. In
  \bibinfo{booktitle}{\emph{The 31st Annual ACM Symposium on User Interface
  Software and Technology Adjunct Proceedings}} \emph{(\bibinfo{series}{UIST
  '18 Adjunct})}. \bibinfo{publisher}{ACM}, \bibinfo{address}{New York, NY,
  USA}, \bibinfo{pages}{29--32}.
\newblock
\showISBNx{978-1-4503-5949-8}
\urldef\tempurl%
\url{https://doi.org/10.1145/3266037.3266091}
\showDOI{\tempurl}


\bibitem[\protect\citeauthoryear{Kovelamudi, Zheng, and Mukkamala}{Kovelamudi
  et~al\mbox{.}}{2016}]%
        {KovelamudiScramble2016}
\bibfield{author}{\bibinfo{person}{Geetika Kovelamudi}, \bibinfo{person}{Jun
  Zheng}, {and} \bibinfo{person}{Srinivas Mukkamala}.}
  \bibinfo{year}{2016}\natexlab{}.
\newblock \showarticletitle{Scramble or not, that is the question a study of
  the security and usability of scramble keypad for {PIN} unlock on
  smartphones}. In \bibinfo{booktitle}{\emph{2016 IEEE/CIC International
  Conference on Communications in China (ICCC)}}. \bibinfo{publisher}{IEEE},
  \bibinfo{address}{Chengdu, China}, \bibinfo{pages}{1--6}.
\newblock
\showISBNx{9781509021437}
\urldef\tempurl%
\url{https://doi.org/10.1109/ICCChina.2016.7636862}
\showDOI{\tempurl}


\bibitem[\protect\citeauthoryear{Kuribara, Shizuki, and Tanaka}{Kuribara
  et~al\mbox{.}}{2014}]%
        {KuribaraRotateAlign2014}
\bibfield{author}{\bibinfo{person}{Takuro Kuribara}, \bibinfo{person}{Buntarou
  Shizuki}, {and} \bibinfo{person}{Jiro Tanaka}.}
  \bibinfo{year}{2014}\natexlab{}.
\newblock \showarticletitle{Vibrainput: Two-step PIN Entry System Based on
  Vibration and Visual Information}. In \bibinfo{booktitle}{\emph{CHI '14
  Extended Abstracts on Human Factors in Computing Systems}}
  \emph{(\bibinfo{series}{CHI EA '14})}. \bibinfo{publisher}{ACM},
  \bibinfo{address}{New York, NY, USA}, \bibinfo{pages}{2473--2478}.
\newblock
\showISBNx{978-1-4503-2474-8}
\urldef\tempurl%
\url{https://doi.org/10.1145/2559206.2581187}
\showDOI{\tempurl}


\bibitem[\protect\citeauthoryear{Kwon and Na}{Kwon and Na}{2014}]%
        {KwonSmudge2014}
\bibfield{author}{\bibinfo{person}{Taekyoung Kwon} {and}
  \bibinfo{person}{Sarang Na}.} \bibinfo{year}{2014}\natexlab{}.
\newblock \showarticletitle{TinyLock: Affordable defense against smudge attacks
  on smartphone pattern lock systems}.
\newblock \bibinfo{journal}{\emph{Computers and Security}}
  \bibinfo{volume}{42} (\bibinfo{year}{2014}), \bibinfo{pages}{137 -- 150}.
\newblock
\showISSN{0167-4048}
\urldef\tempurl%
\url{https://doi.org/10.1016/j.cose.2013.12.001}
\showDOI{\tempurl}


\bibitem[\protect\citeauthoryear{Nguyen, Sae-Bae, and Memon}{Nguyen
  et~al\mbox{.}}{2017}]%
        {NguyenDrawDigit2017}
\bibfield{author}{\bibinfo{person}{Toan~Van Nguyen}, \bibinfo{person}{Napa
  Sae-Bae}, {and} \bibinfo{person}{Nasir Memon}.}
  \bibinfo{year}{2017}\natexlab{}.
\newblock \showarticletitle{DRAW-A-PIN: Authentication using finger-drawn PIN
  on touch devices}.
\newblock \bibinfo{journal}{\emph{Computers \& Security}}  \bibinfo{volume}{66}
  (\bibinfo{year}{2017}), \bibinfo{pages}{115 -- 128}.
\newblock
\showISSN{0167-4048}
\urldef\tempurl%
\url{https://doi.org/10.1016/j.cose.2017.01.008}
\showDOI{\tempurl}


\bibitem[\protect\citeauthoryear{Nyang, Kim, Lee, bae Kang, Cho, Lee, and
  Mohaisen}{Nyang et~al\mbox{.}}{2018}]%
        {NyangTwoThumbs2018}
\bibfield{author}{\bibinfo{person}{DaeHun Nyang}, \bibinfo{person}{Hyoungshick
  Kim}, \bibinfo{person}{Woojoo Lee}, \bibinfo{person}{Sung bae Kang},
  \bibinfo{person}{Geumhwan Cho}, \bibinfo{person}{Mun-Kyu Lee}, {and}
  \bibinfo{person}{Aziz Mohaisen}.} \bibinfo{year}{2018}\natexlab{}.
\newblock \showarticletitle{Two-Thumbs-Up: Physical protection for PIN entry
  secure against recording attacks}.
\newblock \bibinfo{journal}{\emph{Computers \& Security}}  \bibinfo{volume}{78}
  (\bibinfo{year}{2018}), \bibinfo{pages}{1 -- 15}.
\newblock
\showISSN{0167-4048}
\urldef\tempurl%
\url{https://doi.org/10.1016/j.cose.2018.05.012}
\showDOI{\tempurl}


\bibitem[\protect\citeauthoryear{Olade, Liang, and Fleming}{Olade
  et~al\mbox{.}}{2018}]%
        {OladePictures2018}
\bibfield{author}{\bibinfo{person}{Ilesanmi Olade}, \bibinfo{person}{Hai-Ning
  Liang}, {and} \bibinfo{person}{Charles Fleming}.}
  \bibinfo{year}{2018}\natexlab{}.
\newblock \showarticletitle{SemanticLock: An authentication method for mobile
  devices using semantically-linked images}.
\newblock \bibinfo{journal}{\emph{CoRR}}  \bibinfo{volume}{abs/1806.11361}
  (\bibinfo{year}{2018}).
\newblock


\bibitem[\protect\citeauthoryear{Takada and Koike}{Takada and Koike}{2003}]%
        {TakadaAwaseE2003}
\bibfield{author}{\bibinfo{person}{Tetsuji Takada} {and}
  \bibinfo{person}{Hideki Koike}.} \bibinfo{year}{2003}\natexlab{}.
\newblock \showarticletitle{Awase-E: Image-Based Authentication for Mobile
  Phones Using User's Favorite Images}. In
  \bibinfo{booktitle}{\emph{Human-Computer Interaction with Mobile Devices and
  Services}}, \bibfield{editor}{\bibinfo{person}{Luca Chittaro}} (Ed.).
  \bibinfo{publisher}{Springer Berlin Heidelberg}, \bibinfo{address}{Berlin,
  Heidelberg}, \bibinfo{pages}{347--351}.
\newblock


\bibitem[\protect\citeauthoryear{Tan, Keyani, and Czerwinski}{Tan
  et~al\mbox{.}}{2005a}]%
        {ScrambledKeyboardTan2005}
\bibfield{author}{\bibinfo{person}{Desney~S. Tan}, \bibinfo{person}{Pedram
  Keyani}, {and} \bibinfo{person}{Mary Czerwinski}.}
  \bibinfo{year}{2005}\natexlab{a}.
\newblock \showarticletitle{Spy-resistant Keyboard: More Secure Password Entry
  on Public Touch Screen Displays}. In \bibinfo{booktitle}{\emph{Proceedings of
  the 17th Australia Conference on Computer-Human Interaction: Citizens Online:
  Considerations for Today and the Future}} \emph{(\bibinfo{series}{OZCHI
  '05})}. \bibinfo{publisher}{Computer-Human Interaction Special Interest Group
  (CHISIG) of Australia}, \bibinfo{address}{Narrabundah, Australia, Australia},
  \bibinfo{pages}{1--10}.
\newblock
\showISBNx{1-59593-222-4}
\urldef\tempurl%
\url{http://dl.acm.org/citation.cfm?id=1108368.1108393}
\showURL{%
\tempurl}


\bibitem[\protect\citeauthoryear{Tan, Keyani, and Czerwinski}{Tan
  et~al\mbox{.}}{2005b}]%
        {TanMouseDrag2005}
\bibfield{author}{\bibinfo{person}{Desney~S. Tan}, \bibinfo{person}{Pedram
  Keyani}, {and} \bibinfo{person}{Mary Czerwinski}.}
  \bibinfo{year}{2005}\natexlab{b}.
\newblock \showarticletitle{Spy-resistant Keyboard: More Secure Password Entry
  on Public Touch Screen Displays}. In \bibinfo{booktitle}{\emph{Proceedings of
  the 17th Australia Conference on Computer-Human Interaction: Citizens Online:
  Considerations for Today and the Future}} \emph{(\bibinfo{series}{OZCHI
  '05})}. \bibinfo{publisher}{Computer-Human Interaction Special Interest Group
  (CHISIG) of Australia}, \bibinfo{address}{Narrabundah, Australia, Australia},
  \bibinfo{pages}{1--10}.
\newblock
\showISBNx{1-59593-222-4}
\urldef\tempurl%
\url{http://dl.acm.org/citation.cfm?id=1108368.1108393}
\showURL{%
\tempurl}


\bibitem[\protect\citeauthoryear{von Zezschwitz, De~Luca, Brunkow, and
  Hussmann}{von Zezschwitz et~al\mbox{.}}{2015}]%
        {vonZezschwitzSwiPIN2015}
\bibfield{author}{\bibinfo{person}{Emanuel von Zezschwitz},
  \bibinfo{person}{Alexander De~Luca}, \bibinfo{person}{Bruno Brunkow}, {and}
  \bibinfo{person}{Heinrich Hussmann}.} \bibinfo{year}{2015}\natexlab{}.
\newblock \showarticletitle{SwiPIN: Fast and Secure PIN-Entry on Smartphones}.
  In \bibinfo{booktitle}{\emph{Proceedings of the 33rd Annual ACM Conference on
  Human Factors in Computing Systems}} \emph{(\bibinfo{series}{CHI '15})}.
  \bibinfo{publisher}{ACM}, \bibinfo{address}{New York, NY, USA},
  \bibinfo{pages}{1403--1406}.
\newblock
\showISBNx{978-1-4503-3145-6}
\urldef\tempurl%
\url{https://doi.org/10.1145/2702123.2702212}
\showDOI{\tempurl}


\bibitem[\protect\citeauthoryear{Xia, Jota, McCanny, Yu, Forlines, Singh, and
  Wigdor}{Xia et~al\mbox{.}}{2014}]%
        {Xia0Latency2014}
\bibfield{author}{\bibinfo{person}{Haijun Xia}, \bibinfo{person}{Ricardo Jota},
  \bibinfo{person}{Benjamin McCanny}, \bibinfo{person}{Zhe Yu},
  \bibinfo{person}{Clifton Forlines}, \bibinfo{person}{Karan Singh}, {and}
  \bibinfo{person}{Daniel Wigdor}.} \bibinfo{year}{2014}\natexlab{}.
\newblock \showarticletitle{Zero-latency Tapping: Using Hover Information to
  Predict Touch Locations and Eliminate Touchdown Latency}. In
  \bibinfo{booktitle}{\emph{Proceedings of the 27th Annual ACM Symposium on
  User Interface Software and Technology}} \emph{(\bibinfo{series}{UIST '14})}.
  \bibinfo{publisher}{ACM}, \bibinfo{address}{New York, NY, USA},
  \bibinfo{pages}{205--214}.
\newblock
\showISBNx{978-1-4503-3069-5}
\urldef\tempurl%
\url{https://doi.org/10.1145/2642918.2647348}
\showDOI{\tempurl}


\bibitem[\protect\citeauthoryear{Yang, Grossman, Irani, and Fitzmaurice}{Yang
  et~al\mbox{.}}{2011}]%
        {PreTouchTargetSelYang2011}
\bibfield{author}{\bibinfo{person}{Xing-Dong Yang}, \bibinfo{person}{Tovi
  Grossman}, \bibinfo{person}{Pourang Irani}, {and} \bibinfo{person}{George
  Fitzmaurice}.} \bibinfo{year}{2011}\natexlab{}.
\newblock \showarticletitle{TouchCuts and TouchZoom: Enhanced Target Selection
  for Touch Displays Using Finger Proximity Sensing}. In
  \bibinfo{booktitle}{\emph{Proceedings of the SIGCHI Conference on Human
  Factors in Computing Systems}} \emph{(\bibinfo{series}{CHI '11})}.
  \bibinfo{publisher}{ACM}, \bibinfo{address}{New York, NY, USA},
  \bibinfo{pages}{2585--2594}.
\newblock
\showISBNx{978-1-4503-0228-9}
\urldef\tempurl%
\url{https://doi.org/10.1145/1978942.1979319}
\showDOI{\tempurl}


\end{thebibliography}

\end{document}